\documentclass[twocolumn,showpacs,preprintnumbers,amsmath,amssymb,prb]{revtex4}

\usepackage{graphicx}
\usepackage{dcolumn}
\usepackage{bm}
\usepackage{amsmath}
\usepackage[amssymb]{SIunits}

\begin{document}


\title{Aharonov-Bohm effect in a side-gated graphene ring}

\author{M. Huefner \footnote{Both authors have contributed equally to this work.}}
\email{huefner@phys.ethz.ch} \affiliation{Solid State Physics
Laboratory, ETH Zurich, Switzerland}
\author{F. Molitor\footnotemark[1] }
\affiliation{Solid State Physics Laboratory, ETH Zurich,
Switzerland}
\author{A. Jacobsen}
\affiliation{Solid State Physics Laboratory, ETH Zurich,
Switzerland}
\author{A. Pioda}
\affiliation{Solid State Physics Laboratory, ETH Zurich,
Switzerland}
\author{C. Stampfer}
\affiliation{ JARA-FIT and II. Institute of Physics, RWTH Aachen,
Germany}
\author{K. Ensslin}
\affiliation{Solid State Physics Laboratory, ETH Zurich,
Switzerland}
\author{T. Ihn}
\affiliation{Solid State Physics Laboratory, ETH Zurich, Switzerland}

\date{\today}

\begin{abstract}

We investigate the magnetoresistance of a side-gated ring structure
etched out of single-layer graphene. We observe Aharonov-Bohm
oscillations with  about 5\% visibility. We are able to change the
relative phases of the wave functions in the interfering paths and
induce phase jumps of $\pi$ in the Aharonov-Bohm oscillations by
changing the voltage applied to the side gate or the back gate. The
observed data can be interpreted within existing models for 'dirty
metals' giving a phase coherence length of the order of $1\text{
}\micro\meter$ at a temperature of $500\text{ }\milli\kelvin$.

\end{abstract}

\maketitle




The progress in nano-fabrication technology of graphene has lead to
the realization of graphene constrictions
\cite{molitor_condmat2008,stampfer_condmat2008,han_prl2007,chen_physE2007,todd_nanolett2009,liu_condmat2008}
and quantum
dots.\cite{stampfer_apl2008,stampfer_nanolett2008,ponomarenko_science2008,schnez_condmat2008}
The same technology allows one to study phase coherent transport of
charge carriers in single- and multilayer graphene. In
Ref.\,\onlinecite{Graf07}, weak localization and conductance
fluctuations in mesoscopic samples with about seven graphene layers
were investigated. Recent transport measurements on single-layer
pnp- (npn-) junctions created with a narrow top gate were
interpreted in terms of Fabry-Perot
interference.\cite{young_condmat2008} Magnetoconductance
fluctuations and weak localization effects were observed in
single-layers with superconducting
contacts.\cite{miao_science2007,heersche_nature2007} Theoretical
aspects of phase-coherent conductance fluctuations in graphene
nanostructures,\cite{rycerz07} and the Aharonov--Bohm
effect\cite{aharonov_pr59,webb_prl85} have been
addressed.\cite{recher07,rycerz08}

The Aharonov-Bohm effect has been observed in carbon-materials i.e.
carbon nanotubes before \cite{bachtold99,cao04}. Recently the
Aharonov-Bohm effect  was investigated experimentally in
two-terminal graphene ring structures, and a systematic study of its
dependence on temperature, density of charge carriers, and magnetic
field was presented.\cite{russo_prb2008} In this experiment the
visibility of the Aharonov--Bohm oscillations was found to be less
than 1\% at low magnetic fields. It was speculated that this small
value might be due to inhomogeneities in the two interferometer arms
leading to a tunneling constriction that suppressed the
oscillations.

In this paper we present four-terminal magnetotransport through a
side-gated graphene ring of smaller size than the devices studied in
Ref.\,\onlinecite{russo_prb2008}, and demonstrate $h/e$-periodic
Aharonov--Bohm oscillations with a visibility of more than 5\%. In
addition, we demonstrate that a $\pi$-phase shift of the
oscillations can be achieved  by changing the side or back gate
voltages.



\begin{figure}
    \includegraphics[width=0.45\textwidth]{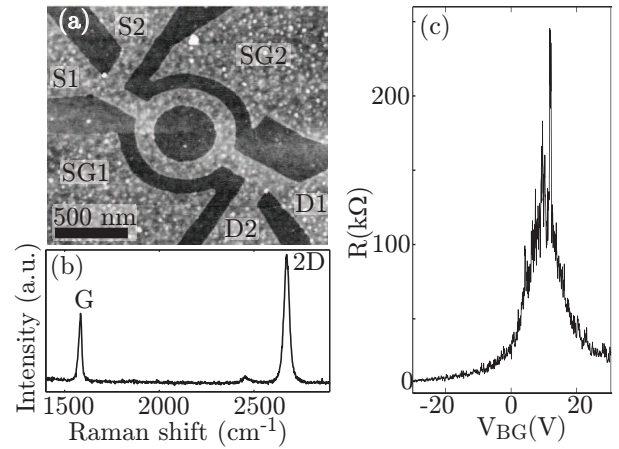}
    \caption{(a) Scanning force micrograph of the ring structure studied in this
    work. The ring has an inner radius of about 200 nm and an outer radius of
    about 350 nm. On each end of the ring structure, there are two graphene
    contact pads labeled S1/2 and D1/2 allowing us to perform four-terminal resistance
    measurements. The side gates labeled SG1 and SG2 are located 100 nm
    away from the structure. (b) Raman spectrum of the same flake before
    processing. The spectrum was recorded using a laser excitation wavelength
    of 532 nm. (c) Four-terminal resistance across the ring structure as a
    function of back gate voltage, with both side gates grounded. The measurement
    is recorded at a temperature of 500~mK with a constant current of 10 nA.}
    \label{fig:fig1}
\end{figure}

Fig. \ref{fig:fig1}(a) displays a scanning force micrograph of the
graphene ring studied in this work. The graphene flakes are produced
by mechanical exfoliation of natural graphite, and deposited on a
highly doped Si wafer covered by $295\text{ }\nano\meter$ of silicon
dioxide.\cite{novoselov_science2004}  Thin flakes are detected by
optical microscopy, and Raman spectroscopy is used to confirm the
single layer nature of the selected graphitic
flakes.\cite{ferrari_prl2006,graf_nanolett2007} In
Fig.\,\ref{fig:fig1}(b) we show the Raman spectrum of the graphene
flake used for the fabrication of the investigated graphene ring
device [Fig.\,\ref{fig:fig1}(a)]. The spectrum has been recorded
before structuring the flake and the
narrow, single Lorentzian shape of the 2D line is evidence for the
single layer nature.\cite{ferrari_prl2006,graf_nanolett2007}
Electron beam lithography, followed by reactive ion etching is used
to define the structure. The contacts are added in a second electron
beam lithography step, followed by the evaporation of Cr/Au (2 nm/40
nm).\cite{stampfer_apl2008}

All measurements presented in this work are performed in a He$^{3}$
cryostat at a base temperature of $T\approx500$ mK. Standard
low-frequency lock-in techniques are used to measure the resistance
by applying a constant current. A magnetic field is applied
perpendicular to the sample plane.





Fig. \ref{fig:fig1}(c) displays the four-terminal resistance of the
ring as a function of applied back gate voltage $V_{\mathrm{BG}}$.
The charge neutrality point occurs at $V_{\mathrm{BG}}\approx
10$\,V. The high resistance observed at the charge neutrality point
is related to the small width $W=150$\,nm of the ring
arms.\cite{molitor_condmat2008} However, this width was chosen large
enough that strong localization of charge carriers leading to
Coulomb-blockade dominated transport in narrow
ribbons\cite{molitor_condmat2008,stampfer_condmat2008} is not
dominant. A rough estimate of the mobility taking into account the
geometry of the structure and using the parallel plate capacitor
model leads to $\mu\approx5000$\,cm$^2/$Vs, comparable to the value
quoted for the material used in Ref.\,\onlinecite{russo_prb2008}.
For the typical back gate voltage $V_{\mathrm{BG}}=-5.8$\,V used for
most of the measurements presented in this paper, the parallel plate
capacitor model gives the sheet carrier density $p_\mathrm{s}=
1.2\times 10^{12}$\,cm$^{-2}$.

We identify the relevant transport regime in terms of appropriate
length scales. The Fermi wavelength corresponding to the carrier
density mentioned above is
$\lambda_\mathrm{F}=\sqrt{4\pi/p_\mathrm{s}}=33$\,nm. For
comparison, at the same density the mean free path is
$l=\hbar\mu\sqrt{\pi p_\mathrm{s}}/e\approx65$\,nm. This is less
than half of the width $W$ of the arms, and much smaller than the
mean ring radius $r_0=275$\,nm and its corresponding circumference
$L=1.7$\,$\mu$m. Therefore, the presented measurements are all close
to the diffusive (dirty metal) regime, and carrier scattering at the
sample boundaries alone cannot fully account for the value of the
mean free path. The relevance of thermal averaging of phase-coherent
effects can be judged from the thermal length $l_\mathrm{th} = \hbar
v_\mathrm{F} l/2k_\mathrm{B}T=700$\,nm, which is significantly
smaller than $L$. This indicates that thermal averaging of
interference contributions to the conductance is expected to be
relevant.



\begin{figure}
    \includegraphics[width=0.45\textwidth]{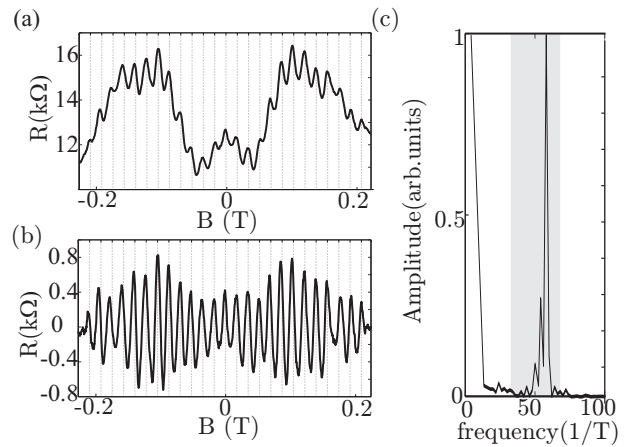}
    \caption{Four-terminal resistance across the ring as a function of magnetic field, recorded at $V_{\mathrm{BG}}$=-5.789 V with a constant current of 5 nA. (a) displays the raw data. For figure (b), the background resistance has been subtracted  as described in the text.(c)
    Fourier transform of the trace.}
    \label{fig:fig2}
\end{figure}

Fig. \ref{fig:fig2}(a) displays the four-terminal resistance of the
ring as a function of magnetic field at $V_{\mathrm{BG}}=-5.789$\,V.
The raw data trace shows a strong modulation of the background
resistance on a magnetic field scale of about 100\,mT. Clear
periodic oscillations can be seen on top of this background. They
have a period in magnetic field $\Delta B_\mathrm{AB}=17.9$\,mT,
indicated by the vertical lines. This period corresponds to the
$h/e$-periodic Aharonov--Bohm oscillations of a ring structure of
271~nm radius, in good agreement with the mean radius $r_0$ of the
ring. \par


Fig. \ref{fig:fig2}(b) shows the same data with the background
resistance subtracted. The background was determined by performing a
running average over one Aharonov--Bohm period $\Delta
B_\mathrm{AB}$. This method was found to lead to no relevant
distortion of the oscillations after background subtraction (with
some exception in Fig. 3 around $B=0\text{ }\tesla$, which is of
minor importance for this paper.) The amplitude of the Aharonov-Bohm
oscillations is modulated as a function of magnetic field on the
same scale as the background resistance, indicating that a finite
number of paths enclosing a range of different areas contribute to
the oscillations. This observation is compatible with the finite
width $W$ of the ring.\cite{washburn86}

In Fig. 2(c) the fast Fourier transform (FFT) of the data in Fig. \ref{fig:fig2}(a) is plotted.
The peak seen at $60\text{ }\milli\tesla$ corresponds
to the $h/e$-periodic Aharonov Bohm effect. The width of this peak is
significantly smaller than the range of frequencies expected from
the range of possible enclosed areas in our geometry (indicated as a
gray shaded region in Fig.\ref{fig:fig2}(c)). We therefore conclude that the paths contributing to the Aharonov-Bohm effect do not cover the entire geometric area of the ring arms.

In this four-terminal measurement, the oscillations have a
visibility of  about 5\%. In general, the observed Aharonov-Bohm
oscillations become more pronounced for smaller current levels, as
expected. The current level of 5\,nA was chosen as a good compromise
between the signal-to-noise ratio of the voltage measurement and the
visibility of the Aharonov--Bohm oscillations.
However, due to limited sample stability, the visibility of the
oscillations at a given back gate voltage depends on the back gate
voltage history. Therefore measurements presented here were taken
only over small ranges of back gate voltage after having allowed the
sample to stabilize in this range.

Higher harmonics, especially  $h/2e$-periodic oscillations, are
neither visible in the magnetoresistance traces, nor do they lead to
a clear peak in the Fourier spectrum (less than 1\% of the
$h/e$-oscillation amplitude). This indicates that the phase
coherence length $l_\varphi<2L$, i.e., it is (significantly) smaller
than twice the circumference of the ring. Given the temperature of
our experiment, this estimate is well compatible
with the phase-coherence lengths reported in Refs.~\onlinecite{Graf07}, \onlinecite{miao_science2007}, \onlinecite{russo_prb2008},
and \onlinecite{Berger06}.

The measurements were taken in a magnetic field range where the
classical cyclotron radius $R_\mathrm{c}=\hbar
k_\mathrm{F}/eB>640$\,nm is bigger than the mean free path $l$, the
ring width $W$, and even the ring diameter. At the same time, Landau
level quantization effects are negligible, because the sample is
studied in the low field regime $\mu B\ll 1$. The only relevant
effect of the magnetic field on the charge carrier dynamics is
therefore caused by the field-induced Aharonov--Bohm phase.

In diffusive ring-shaped systems, conductance fluctuations can
coexist with Aharonov--Bohm oscillations. However, the relevant
magnetic field scale of the conductance fluctuations $\Delta
B_\mathrm{CF}\sim \phi_0/Wl_\varphi$ ($\phi_0=h/e$) can be forced to
be well separated from $\Delta B_\mathrm{AB}=\phi_0/\pi r_0^2$ by
choosing a sufficiently large aspect ratio $r_0/W$. Judging the
situation from the measurement traces in Fig.\,\ref{fig:fig2}(a),
the only candidates for conductance fluctuations are the magnetic
field dependent variations of the background resistance, which occur
on a magnetic field scale that is at least a factor of five larger
than $\Delta B_\mathrm{AB}$. As far as the amplitude of the
modulation of the background can be estimated from
Fig.\,\ref{fig:fig2}(a), it is of the order of the conductance
quantum $e^2/h$ which is reasonable, since the condition
$l_\varphi\sim L$ implies the absence of strong self-averaging over
the ring circumference $L$.



\begin{figure*}
    \includegraphics[width=0.8\textwidth]{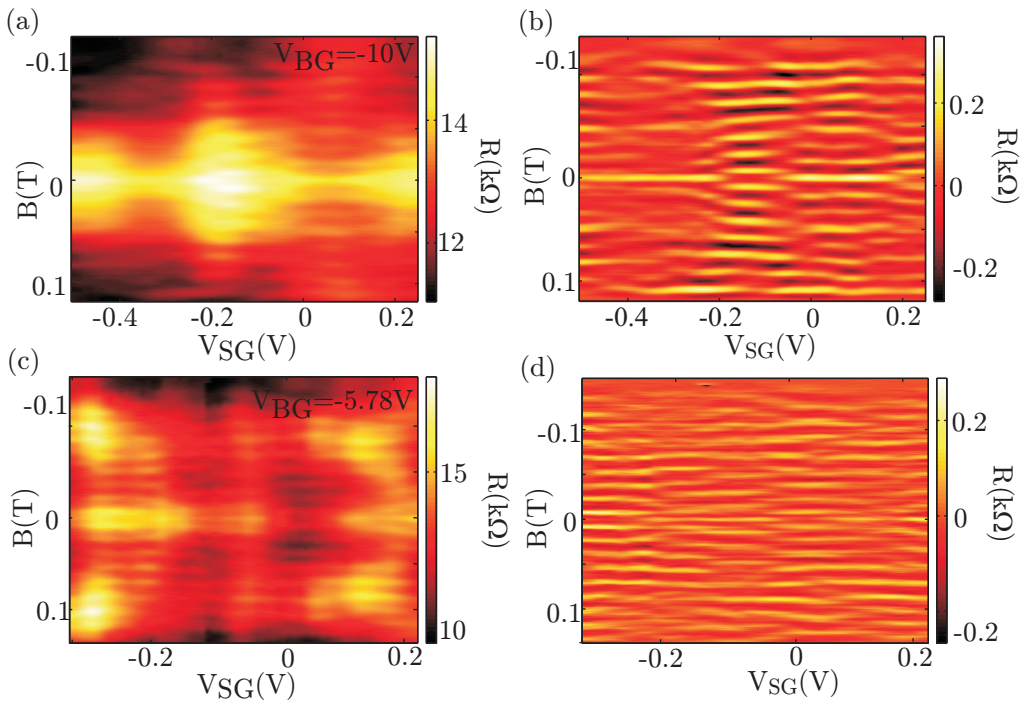}
    \caption{Four-terminal resistance as a function of magnetic field $B$ and voltage applied to SG1 ($V_{\mathrm{SG}}$), measured with a constant current $I=1\text{ }\nano\ampere$. (a) and (c) show the raw data, recorded at $V_{\mathrm{BG}}$=-10~V and $V_{\mathrm{BG}}$=-5.789~V. Figures (b) and (d) show the corresponding data where the background has been removed for each individual trace by averaging over one Aharonov--Bohm period in magnetic field.
}
    \label{fig:fig3}
\end{figure*}

Fig. \ref{fig:fig3} displays the four-terminal resistance of the
ring as a function of magnetic field and voltage $V_{\mathrm{SG}}$
applied to the side gate SG1, for two different back gate voltages
without [Fig.\,\ref{fig:fig3}(a), (c)] and with
[Fig.\,\ref{fig:fig3}(b), (d)] background subtraction.
In the raw data [Fig.\,\ref{fig:fig3}(a), (c)], a modulation of the
background resistance on a magnetic field scale, similar to that in
Fig. \ref{fig:fig2}(a), can be observed. The subtraction of the
background (extracted as described before) makes the Aharonov--Bohm
oscillations visible [Fig.\,\ref{fig:fig3}(b), (d)]. Aharonov--Bohm
oscillations at different $V_\mathrm{SG}$ display either a minimum
or a maximum at $B=0\text{ } \tesla$, with abrupt changes between
the two cases at certain side gate voltage values. This behavior is
compatible with the generalized Onsager symmetry requirement for
two-terminal resistance measurements, $R(B)=R(-B)$. Although our
measurement has been performed in four-terminal configuration, the
contact arrangement with respect to the ring and the fact, that the
contacts are separated by distances $\geq l_{\phi}$ from the ring
lead to a setup where the two terminal symmetry is still very
strong. [c.f., Fig.\,\ref{fig:fig1}(a)]. Closer inspection shows
that the part antisymmetric in magnetic field of each trace (not
shown) is more than a factor of ten smaller than the symmetric
part.\par

In previous studies on metal rings the effect of electric fields on
the Aharonov-Bohm oscillations has been investigated, and two
possible scenarios were discussed:\cite{Washburn87} on one hand, the electric field
may shift electron paths in space and thereby change the
interference. On the other hand, the electric field may change the
electron density and thereby the Fermi-wavelength of the carriers.
We discuss the latter effect in more detail below, since the
relative change in the Fermi wavelength is expected to be more
pronounced in graphene compared to conventional metals.\par

In order to estimate which phase change $\Delta\varphi$ an
electronic wave picks up on the scale of the side gate voltage
change $\Delta V_\mathrm{SG}$ on which Aharonov--Bohm maxima switch
to minima, we use the relation $\Delta\varphi = \Delta
k_\mathrm{F}L_\mathrm{eff}$, where $L_\mathrm{eff}$, being the
effective length of a characteristic diffusive path, is assumed to
be independent of the side gate voltage,\cite{poi} whereas the
change in wave number $\Delta k_\mathrm{F}$ is assumed to be caused
by $\Delta V_\mathrm{SG}$. The quantity $\Delta k_\mathrm{F}$ is
found from the density change $\Delta p_\mathrm{s}$ using $\Delta
k_\mathrm{F} = \sqrt{\pi/4p_\mathrm{s}}\Delta p_\mathrm{s}$. The
density change is related via a parallel plate capacitor model to a
change in back gate voltage, i.e., $\Delta p_\mathrm{s}=\Delta
V_\mathrm{BG}\epsilon\epsilon_0/ed$ ($\epsilon$: relative dielectric
constant of the silicon dioxide substrate, $d$: thickness of the
oxide layer) leading to $\Delta p_\mathrm{s}/\Delta
V_\mathrm{BG}\approx 7.5\times 10^{10}$\,cm$^{-2}$V$^{-1}$. Finally,
$\Delta V_\mathrm{BG}$ is related to $\Delta V_\mathrm{SG}$ via the
lever arm ratio $\alpha_{SG}/\alpha_\mathrm{BG}$.

In order to determine this lever arm ratio, we have performed
measurements of conductance fluctuations in the plane of the back
gate voltage $V_\mathrm{BG}$ and the side gate voltage
$V_\mathrm{SG}$ (not shown). The characteristic slope of fluctuation
minima and maxima in this parameter plane allows us to estimate the
lever arm ratio $\alpha_\mathrm{SG}/\alpha_\mathrm{BG}\approx 0.2$.
In previous experiments on side-gated graphene Hall
bars\cite{molitor_prb2007} we found a similar lever arm for regions
close to the edge of the Hall bar whose width is roughly comparable
to the width of the arms of the ring investigated here.\par

Using the numbers given above and using the density
$p_\mathrm{s}=1.2\times 10^{12}$\,cm$^{-2}$ for
Fig.\,\ref{fig:fig3}(b), we find $\Delta k_\mathrm{F}\approx
1.2\times 10^{6}$\,m$^{-1}$V$^{-1}\Delta V_\mathrm{SG}$. In
ballistic systems the effective length of a path is given by
$L_\mathrm{eff}\sim L$, giving $\Delta\varphi\approx \Delta
V_\mathrm{SG}\pi/1.5$\,V. A phase change of $\pi$ would imply a
change of side gate voltage on the scale of 1.5\,V which is large
compared with the measurement in Fig.\,\ref{fig:fig3}(b) where this
scale is of the order of 100\,mV. However, in the diffusive regime,
a characteristic path contributing to Aharonov--Bohm oscillations is
longer by a factor of $L/l\approx 27$ due to multiple
scattering\cite{imry02} giving $\Delta\varphi\approx \Delta
V_\mathrm{SG}\pi/55$\,mV. A change of the side gate voltage of
typically 55\,mV would cause a switch of the Aharonov--Bohm phase by
$\pi$, in better agreement with the observation than the ballistic
estimate. The same calculation could be used to estimate the
correlation voltage of the conductance fluctuations of the
background resistance, in agreement with the observation in
Fig.\,\ref{fig:fig2} and Fig.\,\ref{fig:fig3}. This correlation
voltage is on the same scale as the phase jumps of the
Aharonov--Bohm oscillations.

\begin{figure*}
    \includegraphics[width=0.8\textwidth]{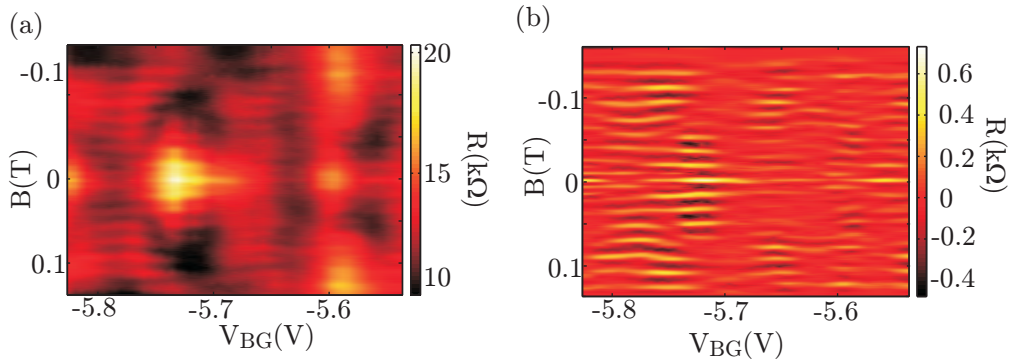}
    \caption{Four-terminal resistance as a function of magnetic field and back gate voltage measured with a constant current of 1~nA. (a) displays the raw data, while for (b) the background has been removed.}
    \label{fig:fig4}
\end{figure*}
Fig.\,\ref{fig:fig4} shows magnetoresistance data for varying back
gate voltages and $V_{\mathrm{SG}}=0\,\mathrm{V}$. Similar to the
case where the side gate was tuned, we observe variations of the
oscillation patterns as well as $\pi$-phase shifts. The raw data
displayed in Fig. \ref{fig:fig4}(a),- shows background fluctuations
with $h/e$-periodic Aharonov--Bohm oscillations superimposed. In
Fig. \ref{fig:fig4}(b), the background has been removed. Again,
alternating minima and maxima at $B=0\text{ } \tesla$ can be
observed.

The larger visibility of Aharonov--Bohm oscillations observed in our
sample, compared to the work in Ref.\,\onlinecite{russo_prb2008} is
unlikely to be caused by better material or sample quality. Also our
measurement temperature is about a factor of four higher than the
lowest temperatures reported there. We therefore believe that the
smaller ring dimensions in combination with the four terminal
arrangement may be responsible for the larger value of the
visibility observed in our experiment. In
Ref.\,\onlinecite{russo_prb2008} the expression \cite{washburn86}
$\Delta G \propto l_\mathrm{th}/l_\varphi\exp(-\pi r_0/l_\varphi)$
was invoked to explain the observed $T^{-1/2}$-dependence of the
oscillation amplitude. The exponential term on the right hand side
contains the radius of the ring $r_0$. A smaller radius will lead to
a larger oscillation amplitude which may explain the improved
amplitude in our measurements. However, trying to relate the
visibilities observed in the two experiments quantitatively
(assuming that all experimental parameters except the ring radius
are the same) would lead to a phase-coherence length $l_\varphi$
smaller than the ring circumference $L$ and only slightly larger
than the ring radius $r_0$. As our experiment demonstrates, a
separation of $h/e$-periodic oscillations from background variations
due to magnetoconductance fluctuations is still possible in our
device despite the aspect ratio $r_0/W$ which is reduced in our
device compared to Ref.\,\onlinecite{russo_prb2008}. A
phase-coherence length between $L$ and $r_0$ is also compatible with
the observation $\Delta B_\mathrm{CF}/\Delta B_\mathrm{AB}\approx
5$.

We also note that the diffusive regime investigated in our device is
quite extended in back gate voltage. Assuming diffusive scattering
at the edges to become dominant as soon as $l\approx W$, we estimate
that this does not occur (for transport in the valence band) until
$V_\mathrm{BG}$ becomes more negative than $-80$\,V. Transport may
also enter a different regime, when the Fermi wavelength becomes
larger than $l$, which is expected to happen (again for transport in
the valence band) at back-gate voltages larger than $+2$\,V in our
sample. An even different regime may be entered at a back gate
voltage of $+9.3$\,V, where $\lambda_\mathrm{F}\approx W$. As a
consequence, the `dirty metal' description of the Aharonov--Bohm
oscillations should be applicable in the whole range of back-gate
voltages shown in Fig.\,\ref{fig:fig1}(c), except for a region of
$\pm 8$\,V around the charge neutrality point, where the resistance
is maximum.


In conclusion, we have observed Aharonov--Bohm oscillations in
four-terminal measurements on a side-gated graphene ring structure.
The visibility of the oscillations is found to be about 5\%. By
changing the voltage applied to the lateral side gate, or the back
gate, we observe phase jumps of $\pi$ compatible with the
generalized Onsager relations for two-terminal measurements. The
observations are in good agreement with an interpretation in terms
of diffusive metallic transport in a ring geometry, and a
phase-coherence length of the order of one micrometer at a
temperature of 500\,mK.


\begin{acknowledgments}
Support by the ETH FIRST lab and financial support by the EH Zuerich
and from the Swiss Science Foundation (Schweizerischer
Nationalfonds, NCCR Nanoscience) are gratefully acknowledged.
\end{acknowledgments}



\begin{thebibliography}{99}
\bibitem{han_prl2007}
    M. Y. Han, B. Ozyilmaz, Y. Zhang, and P. Kim, Phys. Rev. Lett. {\bf 98}, 206805 (2007).


\bibitem{chen_physE2007}
    Z. Chen, Y.-M. Lin, M. J. Rooks, and P. Avouris, Physica E, {\bf 40/2}, 228 (2007).


\bibitem{liu_condmat2008}
X. Liu, J. B. Oostinga, A. F. Morpurgo, L. M.K. Vandersypen, Phys.
Rev. B \textbf{80}, 121407(R), (2009).



\bibitem{todd_nanolett2009}
K. Todd, H. Chou, S. Amasha and D. Goldhaber-Gordon, Nano Lett.,
{\bf 9},1 (2009).


\bibitem{molitor_condmat2008}
F. Molitor, A. Jacobsen, C. Stampfer, J. G\"uttinger, T. Ihn, and K. Ensslin
Phys. Rev. B {\bf 79}, 075426 (2009).



\bibitem{stampfer_condmat2008}
C. Stampfer, J. G\"uttinger, S. Hellm\"uller, F. Molitor, K. Ensslin, and T. Ihn,
Phys. Rev. Lett. {\bf 102}, 056403 (2009).

 \bibitem{ponomarenko_science2008}
  L. A. Ponomarenko, F. Schedin, M. I. Katsnelson, R. Yang, E. H. Hill, K. S. Novoselov, A. K. Geim, Science, {\bf320}, 356 (2008)

\bibitem{stampfer_apl2008}
 C. Stampfer, J. G\"uttinger, F. Molitor, D. Graf, T. Ihn and K. Ensslin, Appl. Phys. Lett {\bf92}, 012102 (2008);

 \bibitem{stampfer_nanolett2008}
 C. Stampfer, E. Schurtenberger, F. Molitor, J. G\"uttinger, T. Ihn and K. Ensslin, Nano. Lett. {\bf8}, 2378 (2008)


\bibitem{schnez_condmat2008}
 S. Schnez, F. Molitor, C. Stampfer, J. G\"uttinger, I. Shorubalko, T. Ihn and K. Ensslin, Appl. Phys. Lett. {\bf94}, 012107 (2009)




\bibitem{Graf07} D. Graf, F. Molitor, T. Ihn, K. Ensslin, Phys. Rev. B {\bf 75}, 245429 (2007).

\bibitem{young_condmat2008}
A. F. Young and P. Kim, Nature Physics {\bf 5}, 222 (2009).

\bibitem{miao_science2007}
F. Miao, S. Wijeratne, Y. Zhang, U. C. Coskun, W. Bao and C. N. Lau, Science {\bf317}, 1530 (2007)

\bibitem{heersche_nature2007}
H. B. Heersche, P. Jarillo-Herrero, J. B. Oostinga, L. M. K. Vandersypen and A. F. Morpurgo, Nature {\bf446}, 56 (2007)

\bibitem{rycerz07} A. Rycerz, J. Tworzydlo, C.W.J. Beenakker, Europhys. Lett. {\bf 79}, 57003 (2007).

\bibitem{aharonov_pr59}
Y. Aharonov and D. Bohm, Phys. Rev. {\bf115}, 485 (1959)

\bibitem{webb_prl85}
R. A. Webb, S. Washburn, C. P. Umbach and R. B. Laibowith, Phys. Rev. Lett. {\bf54}, 2696 (1985)

\bibitem{recher07} P. Recher, B. Trauzettel, A. Rycerz, Ya.M. Blanter, C.W.J. Beenakker, A.F. Morpurgo, Phys. Rev. B {\bf 76}, 235404 (2007).

\bibitem{rycerz08} A. Rycerz, Acta Physica Polonica A {\bf 115}, 322 (2009).

\bibitem{bachtold99} A. Bachtold et al., Nature \textbf{397}, 673
(1999).

\bibitem{cao04} J. Cao et al., Phys. Rev. Lett. \textbf{93}, 216803 (2004).


\bibitem{russo_prb2008}
S. Russo, J. B. Oostinga, D. Wehenkel, H. B. Heersche, S. Shams Sobhani, L. M. K. Vandersypen and A. F. Mopurgo, Phys. Rev. B {\bf77}, 085413 (2008).

 \bibitem{novoselov_science2004}
    K. S. Novoselov, A. K. Geim, S. V. Morozov, D. Jiang, Y. Zhang, S. V. Dubonos, I. V. Grigorieva, and A. A. Firsov, Science {\bf 306}, 666 (2004)

  \bibitem{ferrari_prl2006}
 A. C. Ferrari, J. C. Meyer, V. Scardaci, C. Casiraghi, M. Lazzeri, F. Mauri, S. Piscanec, D. Jiang, K. S. Novoselov, S. Roth, and A. K. Geim, Phys. Rev. Lett. {\bf 97}, 187401 (2006)

\bibitem{graf_nanolett2007}
    D. Graf, F. Molitor, K. Ensslin, C. Stampfer, A. Jungen, C. Hierold, and L. Wirtz, Nano Lett. {\bf 7}, 238 (2007)

\bibitem{onsager_pr31}
L. Onsager, Phys. Rev. {\bf37}, 405 (1931)

\bibitem{onsager_pr31_2}
L. Onsager, Phys. Rev. {\bf38}, 2265 (1931)

\bibitem{buttiker_prl86}
M. B\"uttiker, Phys. Rev. Lett. {\bf 57}, 1761 (1986)




\bibitem{washburn86} S. Washburn, R.A. Webb, Adv. Phys. {\bf 35}, 375 (1986).

\bibitem{Berger06} C. Berger {\it et al.}, Science {\bf 312}, 1191 (2006).

\bibitem{Washburn87} S. Washburn, H. Schmid, D. Kern, and R. A. Webb, Phys. Rev. Lett. {\bf 59}, 1791 (1987); P. G. N. de Vegvar, G. Timp, P. M. Mankiewich, R. Behringer, and J. Cunningham, Phys. Rev. B {\bf 40}, 3491 (1989).

\bibitem{poi} We remark here that this assumption, and the reasoning based on it as given in the main text, corresponds to the usual argument made for dirty metals. However, in graphene the typical length $L_\mathrm{eff}=L^2/l$ of a diffusive path\cite{imry02} is proportional to $k_\mathrm{F}^{-1}$, such that the phase $k_\mathrm{F}L_\mathrm{eff}$ accumulated along such a path is independent of $k_\mathrm{F}$ and therefore independent of carrier density and gate voltage. It therfore remains unclear to us, how the concept of the Thouless energy as an energy scale for wave function correlations can be transferred to the graphene
system.
We nevertheless discuss the estimate based on the assumption of a $k_\mathrm{F}$-independent $L_\mathrm{eff}$ (1) in the absence of any better theory and (2) in accordance with Ref.\,\onlinecite{russo_prb2008}.


\bibitem{molitor_prb2007}
F. Molitor, J. G\"uttinger, C. Stampfer, D. Graf, T. Ihn and K. Ensslin, Phys. Rev. B {\bf 76}, 245426 (2007)


\bibitem{imry02} Y. Imry, {\em Introduction to mesoscopic physics}, 2nd ed., p. 73, Oxford University Press, New York, 2002.



\end{thebibliography}
\end{document}